\documentclass[journal]{new-aiaa}
\usepackage[utf8]{inputenc}

\usepackage{graphicx}
\usepackage{amsmath}
\usepackage[version=4]{mhchem}
\usepackage{siunitx}
\usepackage{longtable,tabularx}
\usepackage{hyperref}
\usepackage{siunitx}
\usepackage{graphicx}
\usepackage{tikz}
\usetikzlibrary{spy}
\usepackage{pgfplots}
\pgfplotsset{compat=newest}
\pgfplotsset{plot coordinates/math parser=false}
\usepackage{xcolor}
\usepackage{pdfpages}
\usetikzlibrary{plotmarks}
\usepackage{todonotes}
\usepackage{caption}
\usepackage{subfigure}

\usepackage{changes}

\newcommand{\Rey}{\textrm{Re}}

\title{Variational Quantum Algorithms for Computational\\ Fluid Dynamics}

\author{Dieter Jaksch\footnote{Professor of Physics, Institute for Laser Physics, dieter.jaksch@uni-hamburg.de}}
\affil{Universität Hamburg, Luruper Chaussee 149, Gebäude 69, D-22761 Hamburg, Germany}
\affil{Clarendon Laboratory, University of Oxford, Parks Road, Oxford OX1 3PU, UK}
\author{Peyman Givi \footnote{Professor, Mechanical Egineering and Petroleum Engineering, Fellow, AIAA, pgivi@pitt.edu.}}
\affil{University of Pittsburgh, Pittsburgh, Pennsylvania 15260, USA} 
\author{Andrew J.\ Daley \footnote{Professor, Department of Physics and SUPA, andrew.daley@strath.ac.uk.}}
\affil{University of Strathclyde, Glasgow, Scotland G4 0NG, United Kingdom}
\author{Thomas Rung \footnote{Professor, Institute for Fluid Dynamics and Ship Theory, thomas.rung@tuhh.de.}}
\affil{Hamburg University of Technology, Am Schwarzenberg-Campus 4, D-21073 Hamburg, Germany}

\begin{document}
\newlength\figureheight
\newlength\figurewidth

\maketitle
\begin{abstract}
Quantum computing uses the physical principles of very small systems to develop computing platforms which can solve problems that are intractable on conventional supercomputers. There are challenges not only in building the required hardware, but also in identifying the most promising application areas and developing the corresponding quantum algorithms. The availability of intermediate-scale noisy quantum computers is now propelling the developments of novel algorithms, with applications across a variety of domains, including in aeroscience. Variational quantum algorithms are particularly promising since they are comparatively noise tolerant and aim to achieve a quantum advantage with only a few hundred qubits. Furthermore, they are applicable to a wide range of optimization problems arising throughout the natural sciences and industry. To demonstrate the possibilities for the aeroscience community, we give a perspective on how variational quantum algorithms can be utilized in computational fluid dynamics. We discuss how classical problems are translated into quantum algorithms and their logarithmic scaling with problem size. As an explicit example we apply this method to Burgers’ Equation in one spatial dimension. We argue that a quantum advantage over classical computing methods could be achieved by the end of this decade if quantum hardware progresses as currently envisaged and emphasize the importance of joining up development of quantum algorithms with application-specific expertise to achieve real-world impact.
\end{abstract}
%
%

\section{Introduction 
}
\label{sec:into}
\subsection{Quantum Computing}
Progress in quantum computing is driven  by its potential to revolutionize our ability to solve challenging computational problems that are otherwise intractable. We expect quantum processors to soon be integrated into existing high performance computing (HPC) infrastructure, much as Graphics Processing Units (GPUs) are today. Because the underlying quantum physical principles are entirely different from conventional computers, this is not business-as-usual for the development of applications and algorithms, which must be built from the ground up. It is a challenging process, which requires knowledge of quantum computing and expertise in the problem itself. 
Seminal, but niche, quantum algorithms requiring high-quality hardware e.g. for factoring numbers \cite{Shor_97} and searching unstructured databases \cite{Grover_1996}  have existed for decades. More recently, discoveries have been made that may yield quantum advantage in solving a wide range of numerical problems, including numerous industrially relevant ones \cite{Montanaro2016}. Quantum algorithms have the potential to outperform classical computation in simulation and optimization targeting several industries including pharmacology, chemistry, energy, financial, aeroscience, insurance and logistics. A nascent quantum technologies market has been developed on this basis \cite{Bobier_2021}. For instance, a variety of applications are envisaged for aeroscience and engineering  specifically, and there has been substantial work on the potential impact of quantum computing in this sector (see Ref.\ \cite{Givi_2020} for a review). A quantum advantage over classical algorithms has been found for problems designed to test the quantum hardware \cite{Arute_2019}, and special-purpose quantum simulators are already having an impact on quantum science \cite{Daley2022}. However, current technology is still limited to comparatively small and noisy systems characterizing the noisy intermediate-scale quantum (NISQ) era. In this era, quantum algorithms that are forgiving to errors and may give a quantum advantage over conventional computing using only hundreds of qubits are sought. 

In this article, we illustrate the opportunities for aeroscience and engineering, focusing on variational algorithms, and specifically on applications in computational fluid dynamics. This is a good example of how we can connect the very different approaches to computational challenges implied by quantum computers to a specific problem of practical interest in the aeroscience community. We describe the background of the algorithm and details of possible applications, before coming back to a broader perspective on the future of these approaches.

\subsection{Variational Quantum Algorithms (VQAs)}
The power of quantum computing arises from the possibility of the system to be in a superposition of multiple states simultaneously. Where in a conventional classical computer information is stored as bits, which can be $0$ or $1$, in a quantum computer we instead have qubits, which can take any superposition of $0$ and $1$, with complex coefficients. To describe the state of $N$ bits in a classical computer, we need to specify values of $0$ or $1$ for each of the bits. In contrast, describing the wave function of $N$ qubits requires $2^N-1$ complex coefficients. This gives an exponentially growing capacity to store information, but does not immediately mean that we can run existing algorithms faster. The essential challenge is that whenever we measure a quantum state, we only obtain $N$ bits of information – so we do not have direct access to the information stored in the whole state. As a result, we need to find new ways to perform calculations that make use of this large state space, but 
only require restricted numbers of input and output parameters. This can e.g.~be achieved by hybrid algorithms that use classical high performance computing in tandem with quantum computing for specific subroutines.

A promising hybrid quantum-classical algorithm \cite{Peruzzo_2014} was invented in 2014. This algorithm distributes the tasks of solving an optimization problem between a quantum computer that takes as its inputs classical variational parameters $\boldsymbol \lambda$ and calculates a cost function as its output, and a conventional classical computer taking cost function values as its input to update the variational parameters. The two parts work together iteratively to optimize the cost function, which e.g. represents the  residuals of partial differential equations (PDE). The resulting variational parameters $\boldsymbol \lambda_{\rm opt}$ comprise a classical description of the optimal trial solution. Note that this trial solution can usually only be worked out and be analyzed efficiently on the quantum device, making use of the exponentially large state-space, but we only need a set of classical input parameters, and an output that is not the state itself, but only the measured value of a cost function – thus making use of the large state space, but not needing to read it out.

This approach to quantum computing has been rapidly developed over the past years, and has several beneficial properties for NISQ era devices \cite{Cerezo2021}. First, no quantum memory is required for storing the solution, since its description is contained in the classical parameters $\boldsymbol \lambda$. Second, cost function values can often be read out via a single qubit (other schemes are possible) thus requiring only measurements on a single qubit. Third, several error mitigation strategies have been developed (for early examples see \cite{McArdle_2019}) for digital simulation algorithms that are suitable for NISQ devices. Fourth, variational quantum algorithms can be applied to a wide range of linear optimization problems with relevance extending far beyond the natural sciences.
An extension of variational quantum algorithms to nonlinear variational problems has been proposed \cite{Lubasch_2020}. At its heart, this scheme contains a quantum nonlinear processing unit (QNPU) that evaluates cost functions that are polynomials of individual trial functions and can thus be applied to a wide range of nonlinear problems including nonlinear PDEs.
\subsection{Applications of Variational Quantum Algorithms in  computational fluid dynamics (CFD)}
\label{sec:into-appl}
Because of their demonstrated capabilities, VQAs are expected to have a major influence in aeroscience and engineering research. With their success in approaching non-linear differential equations comes the opportunity to apply these algorithms to solve a wide class of problems in transport phenomena – in particular fluid dynamics, which is the primary focus of this article. CFD is widely regarded as one of the major constituents of aerospace research, with applications in a large variety of other disciplines. For diagnostic purposes, i.e. for analyzing existing flow configurations, powerful CFD tools are widely available. These tools typically use approximate physical models to represent unresolved phenomena and employ scale reducing simplifications and constraints to increase the realizability of the computations. Difficulties in identifying broadly applicable physical models and approximations severely limit the usage and general reliability of CFD methods for predictive and design purposes.

For aeroscience and engineering applications, the strengths and weaknesses of CFD are most clearly evident in turbulent flow modeling and simulation. Dealing with the wide range of spatial and temporal scales in such flows remains a grand CFD challenge. The complexity is substantially escalated in high speed and/or reactive flows as occurring in aero \& propulsion systems. In these flows, the coupling of turbulence with compressibility and chemistry makes CFD implementations significantly more difficult. The NASA CFD vision 2030 \cite{Slotnick_2014, Cary_2022}  concludes that present CFD capabilities must undergo a revolutionary change to meet future needs of academia and industry.

Turbulent flow simulations can be classified into two categories\cite{Givi_1989}: (i) direct, and (ii) modeled. In the first, typically termed direct numerical simulation (DNS), the fundamental transport equations are solved in a brute force manner with an attempt to resolve all of the flow scales explicitly. In the second, these equations are “averaged” in some fashion to decrease the range of the scales for a more affordable computation. Reynolds averaged Navier-Stokes (RANS) \cite{Wilcox_2006} and large eddy simulation (LES) \cite{Sagaut_2010} are examples of such simulations. In RANS, the equations are ensemble-averaged at all spatial scales. In LES, the averaging is conducted by pre-filtering of the equations with a certain spatial filter. The resulting modelled simulations are generally conduced in a deterministic or probabilistic manner.   In the former,  finite moments of the unresolved fields are modeled via algebraic or differential equations. In the latter, these fields are described stochastically, effectively by a Fokker-Planck equation. The probabilistic closures in RANS and LES are referred to as the probability density function (PDF) \cite{Pope_1985} and the filtered density function (FDF) \cite{Givi_2006}, respectively.

The fidelity of modeled simulations is directly hinged upon the success of its constituent closures constructed to capture the essential physics. The ultimate fidelity, achievable only via DNS, can be realized in simplistic applications of transport with low to moderate Reynolds numbers. This is particularly the case for multi-physics applications such as conjugate heat transfer, compressible and chemically reacting flows, where the added physical complexity further compounds the problem. Using DNS for analyzing and optimizing technical flows is the holy grail of CFD research and would enable reliable design of much more energy efficient products and infrastructure.

Quantum CFD (QCFD) promises to fundamentally change this paradigm by stripping away the need for limits, simplification and specialization, through the direct resolution of all relevant physical scales and phenomena by encoding flow fields in an exponentially large Hilbert space. More importantly, it promises to do so for a small fraction of the computational cost required by classical methods. The impact of such a technology on the field of CFD, in particular, and product design, in general, would be nothing short of revolutionary. More generally, it would massively strengthen numerical simulations and data sciences as the third pillar of today’s scientific enterprise next to theory and physical experiments.

Various approaches to CFD on quantum computers have recently been put forward by the research community. These have mostly concentrated on efficient methods for calculating the dynamical evolution of flows. For instance, proposal \cite{Lloyd_2020} is based on the derivation of the nonlinear Schrödinger equation from the linear Schrödinger equation for quantum many-body systems, i.e. constitutes a mean-field approximation, which is accurate in the limit of weak interactions. The approach \cite{Liu_2021} uses the technique of Carleman linearization to map a specific set of weakly nonlinear ordinary differential equations with dissipation to a higher-dimensional linear problem. Both methods then solve linear problems using the quantum linear systems algorithm \cite{Harrow_2009}. This allows them to be efficient in the number of time steps and leads to mathematically rigorous convergence guarantees for weak nonlinearities. Furthermore, machine learning approaches towards solving nonlinear partial differential equations have also been proposed for use in CFD \cite{Kyriienko_2021}.

In this perspective we focus our discussion on utilizing the QNPU extension \cite{Lubasch_2020} of the widely used standard VQA \cite{Peruzzo_2014} for solving nonlinear partial differential equations. The QNPU approach uses multigrid renormalization \cite{Lubasch_2018} to encode trial functions in coordinate space described by $M=2^N$ discrete values into $N$ qubits. The size of the required quantum registers thus scales logarithmically with the discrete size of the spatial grid and hence the range of length scales that can to be covered in the CFD problem. Quantum networks implementing problem specific cost function evaluations inside the QNPU are programmed using Matrix Product Operator methods \cite{Schollwock2011densitymatrix} as a high-level quantum programming paradigm \cite{Lubasch_2020}. This method shares with \cite{Lloyd_2020,Kyriienko_2021} that it requires a number of qubits that scales logarithmically with the number of spatial grid points but is not limited to small nonlinearities. QCFD algorithms based on this logarithmic encoding can still only gain a quantum advantage if shallow variational networks creating a subset of trial functions are sufficiently expressive to yield accurate approximations to the actual solution. 

Because of this restricted set of trial functions, VQA can be viewed as a {\it structure} resolving scheme and be classified as a reduced order model (ROM), in which only a  part of Hilbert space is needed to capture the essential flow physics. In this aspect, the methodology portrays some similarity to, albeit being fundamentally different from, the currently popular CFD-ROM strategies \cite{Kutz2016,RNB21}. Most current ROMs are {\it data-driven}, whereas VQA is a fully physics based methodology. The multi-grid encoding of turbulence structures through scale correlations into entangled quantum states of qubits yields shallow quantum variational networks defined by a small number of variational parameters. In Ref.~\cite{Gourianov_2022, gourianov2022} it is shown that such VQA simulations outperform under-resolved DNS (or LES with no sub-grid scale closure) of the flow with the same number of computational degrees of freedom.

As shown in \cite{Gourianov_2022} and discussed in this article, this enables variational QCFD (VQCFD) algorithms to scale at least polynomially better with the Reynolds number $\Rey$ than corresponding classical Navier-Stokes simulations for paradigmatic flow examples in two and three spatial dimensions (an exponential quantum speedup may also be possible but has not been proven). Here, $\Rey$ describes the ratio between the largest and smallest length scales that need to be resolved by the simulation. VQCFD algorithms thus gain quantum advantages in the number of required qubits and algorithm scaling.

We note that the connection between VQCFD algorithms, matrix product operators and matrix product states (MPS) \cite{Schollwock2011densitymatrix} may inspire novel classical numerical tools for solving CFD problems \cite{Gourianov_2022, gourianov2022}. Recent progress in highly efficient numerical tensor manipulations on classical tensor-processing-units \cite{Ganahl2022} demonstrates the potential of classical tensor network methods to compete with NISQ devices and more traditional high-performance classical computing approaches. A detailed discussion of such quantum inspired approaches to CFD is beyond the scope of the current perspective.
\section{The VQCFD Methodology}
\label{sec:secMETH}
Intrinsically, the QNPU approach is an optimization method where cost functions are calculated from parameterized trial solutions to non-linear PDEs on a quantum processor. Boundary and design conditions are encoded in the specific form of the PDEs. Optimization proceeds by iteratively improving the parameters of the trial solutions using classical optimization methods and quantum mechanically evaluating the updated cost function. Application of this approach to steady and unsteady CFD problems requires rewriting PDEs in terms of optimization problems. Importantly, this allows treating physical constraints on the flow properties and design constraints at the same level and optimizing both simultaneously.

\begin{figure}[hbt!]
\centering
\includegraphics[width=.4\textwidth]{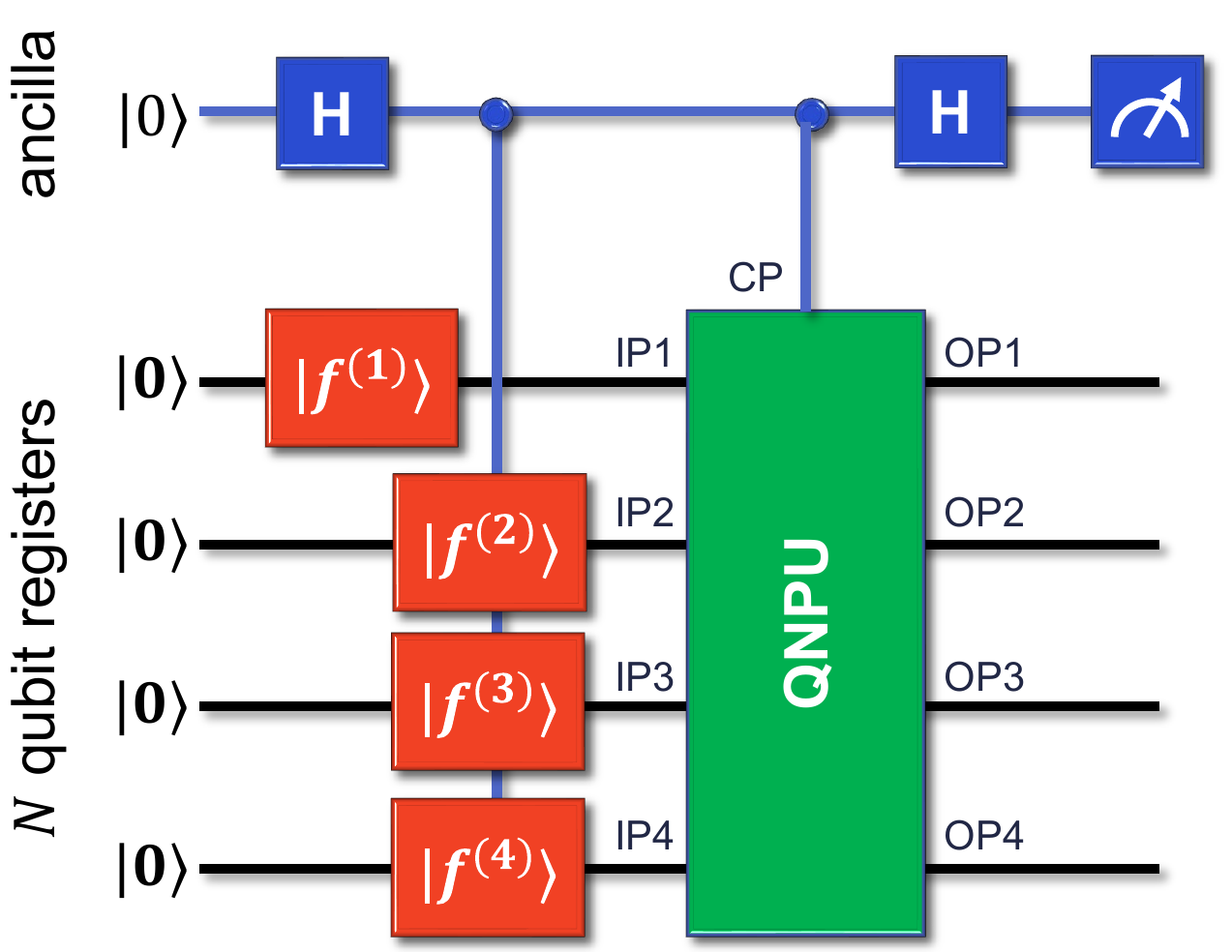}
\caption{
The QNPU architecture for four trial states 
$|f^{(j)} \rangle$ on four $N$ qubit quantum registers initialized to $|\boldsymbol 0\rangle=|00 \dots 0 \rangle$. The red parts of the network create variational trial states. The green QNPU part implements the problem specific linear operators $O_j$. Its operation is controlled by port CP, trial functions enter through input ports IPx, and outputs are labelled OPx. The blue ancilla network is used to evaluate the cost function (figure reproduced from \cite{Lubasch_2020}).
}
\label{fig:fig1}
\end{figure}

The hybrid quantum-classical method for non-linear optimization that forms the basis of VQCFD algorithms was introduced by Lubasch et al. in \cite{Lubasch_2020} and is described in the remainder of this section. The quantum part of the generic algorithm is displayed in Figure \ref{fig:fig1}. It proceeds by encoding trial solutions, generically labeled $f^{(i)}$ here, representing flow fields and also flow design parameters into quantum states $|f^{(i)}\rangle$ created by variational quantum networks, (cf. Sec. \ref{subsec:multigrid}). The network gates are controlled by a set of classical variational parameters denoted by the vector $\boldsymbol \lambda$, as described in Sec. \ref{subsec:lambda}. The problem specific non-linear cost function is evaluated from these trial quantum states in the QNPU, see Sec. \ref{subsec:QNPU}. A unique feature of the QNPU architecture is that it makes use of multiple copies of a variational state $|f^{(i)} \rangle =|f^{(j)}\rangle$ for $i \ne j$ to create powers of a trial solution and hence to realize nonlinear terms in the cost function. The QNPU architecture is therefore not limited to solving only weakly nonlinear problems, which is deemed to be a major advantage. The general form of functions $F$ that can efficiently be generated in this way is given by
\begin{equation}
     F=f^{(1)*}  \Pi_{j=1}^r (O_j f^{(j)}).
     \label{eq:FGENERATION}
\end{equation}
Here, $O_j$ are linear differential operators acting on $f^{(j)}$. These operators are problem specific and implemented in the QNPU. The operators $O_j$ are programmed utilizing Matrix-Operator methods as a high-level programming paradigm \cite{Lubasch_2020}. The cost function, measured via an ancilla qubit shown in Figure \ref{fig:fig1}, is then given by the sum over the real parts of function values of $F$, i.e. $\mathcal C=\sum_k \mathcal R\{F_k\}$, as addressed in Sec. \ref{subsec:CMEASURE}. Based on the measured cost function values, the variational parameters $\boldsymbol \lambda$ are iteratively updated on classical computing hardware until the minimum of the cost function is found using well-established VQA optimization methods \cite{Peruzzo_2014}. Importantly, this setup does not require difficult to realize long-term quantum memory, all relevant problem parameters are stored classically in $\boldsymbol \lambda$.

 \subsection{Trial function encoding} 
 \label{subsec:multigrid}
\begin{figure}[hbt!]
\centering
\includegraphics[width=.4\textwidth]{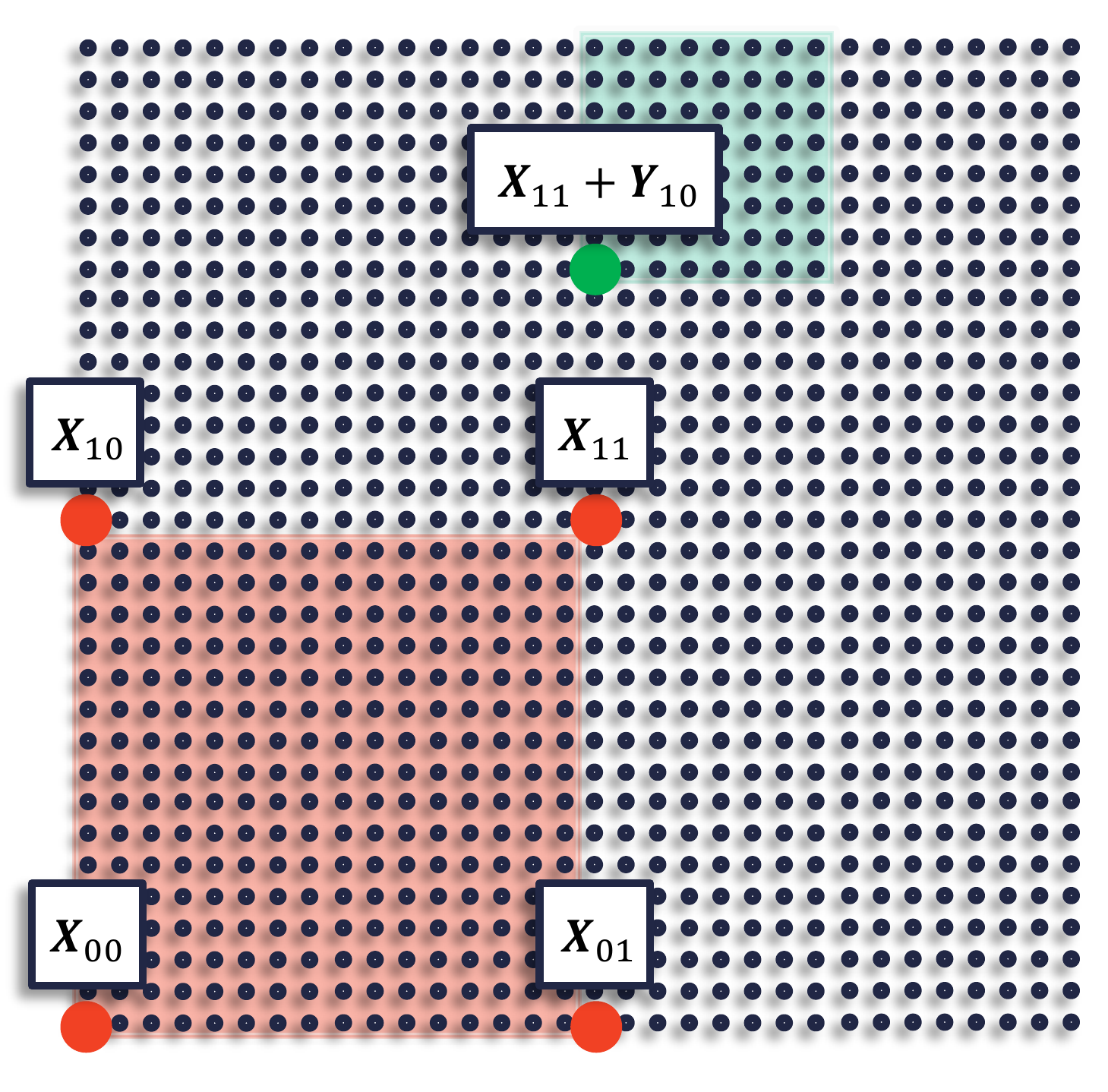}
\caption{
 A cartesian two dimensional $2^5 \times 2^5$ grid illustrating the hierarchical multigrid decomposition. Red circles show the points $\boldsymbol X_{kl}$ of the coarsest 
 $2 \times 2$ grid. One of the four  $2^4 \times 2^4$ sub-grids is indicated by a red-shaded square. The green circle shows a grid point on one of the $2^4 \times 2^4$ sub-grids and the green-shaded square covers one of the 16 sub-grids of size $2^3 \times 2^3$.
}
\label{fig:fig2}
\end{figure}

The trial functions typically represent spatially varying properties of flows or design constraints on a discrete mesh. In QCFD algorithms a function defined by $2^N$ (complex) values is stored as the probability amplitudes of an $N$ qubit register so that quantum register requirements only grow logarithmically with the discrete size of the CFD problem. Moving to increasingly fine meshes is thus far less memory demanding than in standard CFD computations. The drawback of this encoding is that, in general, the complexity of the variational quantum network (defined by classical parameters $\boldsymbol \lambda$) required to create sufficiently expressive trial functions is not known. Large amounts of entanglement and deep variational quantum networks might be required thus limiting the possible quantum advantage. For structured meshes a hierarchical multigrid encoding of trial functions has been developed that makes it possible to exploit the scale locality of fluid flows to limit the depth of the required variational quantum networks \cite{Lubasch_2018}. The basic idea is to iteratively divide the mesh into a series of sub-meshes describing increasingly local features of the flow and illustrated in Figure \ref{fig:fig2} for a two-dimensional $2^{N'} \times 2^{N'}$ grid with points $\boldsymbol r_{ij}$, where $N=2N'$.

The $2^N$ normalized function values are encoded as the probability amplitudes of the wave function of $N$ qubits. The decomposition starts by dividing the grid into four sub-grids of size $2^{(N'-1)} \times 2^{(N'-1)}$. Using a Schmidt decomposition, one correspondingly writes the function values $f(\boldsymbol r_{ij})$ as a sum
              \begin{equation}
                  f(r_{ij})= 
                  \sum_{\alpha=1}^4\lambda_\alpha
                   R_\alpha ({\boldsymbol X_{kl}} ) f_\alpha ({\boldsymbol x_{mn}})
                   \label{eq:scedomp1}
              \end{equation}    
 where $\boldsymbol r_{ij}=\boldsymbol X_{kl}+\boldsymbol x_{mn}$ with $\boldsymbol X_{kl}$ 
 ($0\le k$, $l<2$) a point on a coarse $2 \times 2$ grid and $\boldsymbol x_{mn}$ ($0\le m$, $n<2^{N'-1}$) a 
point on the fine $2^{N'-1} \times 2^{N'-1}$ grid (cf. Figure \ref{fig:fig2}). The Schmidt decomposition gives orthogonal functions $R_\alpha$ and $f_\alpha$ fulfilling
\begin{equation}
    \sum_{kl} R_\alpha (\boldsymbol X_{kl} ) R_\beta (\boldsymbol X_{kl} )= \sum_{mn}f_\alpha (\boldsymbol x_{mn} ) f_\beta (\boldsymbol x_{mn} ) =\delta_{\alpha \beta}
\end{equation}
and real positive values $\lambda_\alpha$  that that obey $\sum_\alpha \lambda_\alpha=1$. The Schmidt values are ordered as $\lambda_1 \ge \lambda_2 \ge \dots \ge \lambda_4 \ge 0$ and provide a measure of the weight of the corresponding term in the sum. Note that at most four terms are required in the sum to represent all possible functions since the decomposition splits off four grid points ${\boldsymbol X_{kl}}$. Truncating the sum by dropping terms $l$ with small $\lambda_l$ will still give a good approximation to the originally encoded function. 

The four function values $R_\alpha (\boldsymbol X_{kl})$ are written into the first two qubits of the quantum register and represent the coarsest features of the function $f$. The qubit wave function encoding $f$ is thus written as 
\begin{equation}
    |\psi\rangle= \sum_\alpha \lambda_\alpha \,  |R_\alpha \rangle \, |f_\alpha \rangle
    \label{eq:wavefunc1}
\end{equation}
where $|R_\alpha \rangle =\sum_{kl} R_\alpha (\boldsymbol X_{kl} )|kl\rangle$ and similarly for $|f_\alpha \rangle$. Here $|kl\rangle \in \{|00\rangle,|01\rangle,|10\rangle,|11\rangle\}$ are the computational basis states of the first two qubits.
The decomposition of the functions $f_\alpha$ proceeds by dividing the $2^{N'-1} \times 2^{N'-1}$ grids into four grids of size $2^{N'-2} \times 2^{N'-2}$ each. The next iteration of the decomposition then reads
\begin{equation}
    f(r_{ij} )= \sum_{\alpha=1}^4 \lambda_\alpha  R_\alpha (\boldsymbol X_{kl} ) \sum_{\beta=1}^4 \lambda_{\alpha\beta}  R_{\alpha\beta} (\boldsymbol Y_{kl} ) f_{\alpha\beta} (\boldsymbol y_{op})
     \label{eq:scedomp2}
\end{equation}
with $\boldsymbol x_{mn}=\boldsymbol Y_{kl}+\boldsymbol y_{op}$ where $\boldsymbol Y_{kl} (0\le k, l<2)$ and $\boldsymbol y_{op} (0\le m, n<2^{N'-2})$ are points on coarse and fine grids, respectively. As in the previous step, one again writes the corresponding wave function as
\begin{equation}
    |\psi\rangle=\sum_{\alpha}\lambda_\alpha |R_\alpha \rangle \, \sum_\beta \lambda_{\alpha\beta}  |R_{\alpha\beta} \rangle \, |f_{\alpha\beta}\rangle 
    \label{eq:wavefunc2}
\end{equation}
with the coarse features of $f_\alpha$ encoded in the second and third qubits of the register 
$|R_{\alpha\beta} \rangle =\sum_{kl} R_{\alpha\beta} (\boldsymbol Y_{kl} ) |kl\rangle$. This procedure is repeated until the fine grid is of size $2 \times 2$.

This is identical to the MPS decomposition of a many-body wave function in one spatial dimension frequently used when simulating many-body quantum systems \cite{Schollwock2011densitymatrix}. However, neighboring sites of the quantum many-body lattice system are here replaced by by neighboring length scales of the mesh \cite{Gourianov_2022}. 

This mapping possesses several features that are important for simulating fluid flow configurations. First, the significance of the qubits in the quantum register corresponds to the length scales of the flow features that they describe. The most significant bits store information about the largest features of the flow while the least significant qubits represent the finest features appearing in the flow. Tracing out more and more of the least significant qubits yields an increasingly coarse representation of the original function with the fine details being averaged out. The number of qubits required to cover the range of all relevant length scales from the largest size of the energy containing eddies $l$, down to the smallest ones, known as the Kolmogorov microscale $\eta$ increases only logarithmically with the ratio $l/\eta$. Specifically, in $K$ spatial dimensions it is given by $N=K \log_2 l/\eta$ in contrast to standard grid based CFD methods where memory resources tend to scale like $l/\eta$. The separation between these length scales often goes like $l/\eta =\Rey^{3/4}$ in each spatial dimension giving an estimate for the number of qubits required to hold a scalar function $f$ in $K$ dimensions as
\begin{equation}
    N=\frac{3K}{4}  \log_2 \Rey \,.
    \label{eq:scaling}
\end{equation}

\begin{figure}[hbt!]
\centering
\includegraphics[width=.4\textwidth]{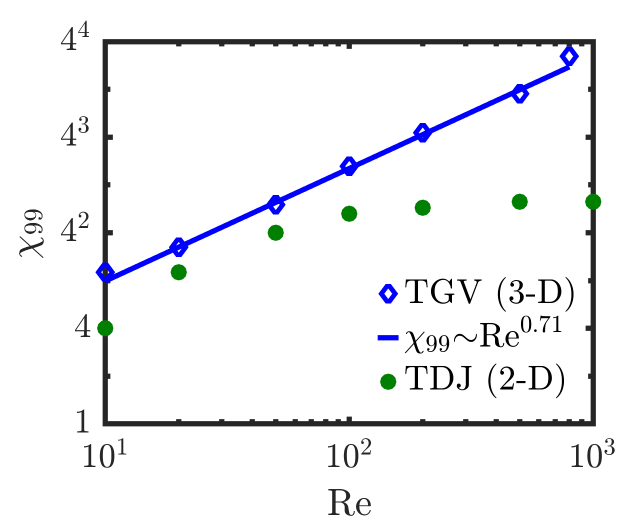}
\caption{
 The Schmidt number $\chi_{99}$, which is a measure of the matrix size required in a 99\% accurate MPS representation of a flow field, as a function of $\Rey$. The blue diamonds refer to a decaying Taylor-Green vortex (TGV) in 3D where $\chi_{99} \propto \Rey^{0.71}$ for sufficiently large Re. The green dots arise from a 2D temporally developing jet (TDJ) simulation, where $\chi_{99} \propto \Rey^0$, i.e. the Schmidt number decouples from the Reynolds number for sufficiently large $\Rey$. Figure taken from \cite{Gourianov_2022}.}
\label{fig:fig3}
\end{figure}

Second, the entanglement between qubits describes correlations of flow features present at different length scales, and not correlations between different regions of the flow. If no correlations exist between length scales the trial functions f are described by a product state. Only in cases where strong correlations exist between all length scales will the maximum amount of entanglement between qubits be required to get an accurate description of a function $f$. For a grid that resolves space down to the microscale $\eta$ the entanglement entropy between the first Kn and the last $K(N-n)$ qubits is a measure for the interscale correlations between length scales larger and smaller than $2^{N-n} \eta$. The multigrid encoding is thus highly suited to exploiting the scale locality of flow physics, which we will here discuss for the turbulent energy cascade \cite{Kolmogorov_1941, Chen_2006, Eyink_2005, Cardesa_2017}. This structure implies that significant correlations only exist between nearby length scales, while very little correlation is expected between the largest and the smallest length scales of a flow. A quantitative analysis of flow examples in \cite{Gourianov_2022} reveals that this lack of correlations between far separated length scales translates into small amounts of entanglement, i.e. only a small number of terms in the above multigrid encoding are significant, while most can be neglected.
This was shown in \cite{Gourianov_2022, gourianov2022}  by decomposing numerically exact solutions to paradigm fluid flows into MPS and investigating the loss of fidelity when truncating the number of terms $\chi$ – also known as the Schmidt number or bond dimension of a MPS – allowed in the sums representing $|\psi\rangle$. The results of this analysis are reproduced in Figure \ref{fig:fig3} which shows that $\chi_{99}$, required for a truncated MPS state being 99\% accurate, to only grow polynomially with the Reynolds number for a temporally developing jet (TDJ) in 2D and a decaying Taylor-Green vortex (TGV) in 3D. As discussed next, variational quantum trial functions can thus be created in shallow quantum networks.

\subsection{Creating variational trial states}
 \label{subsec:lambda}
\begin{figure}[hbt!]
\centering
\includegraphics[width=.4\textwidth]{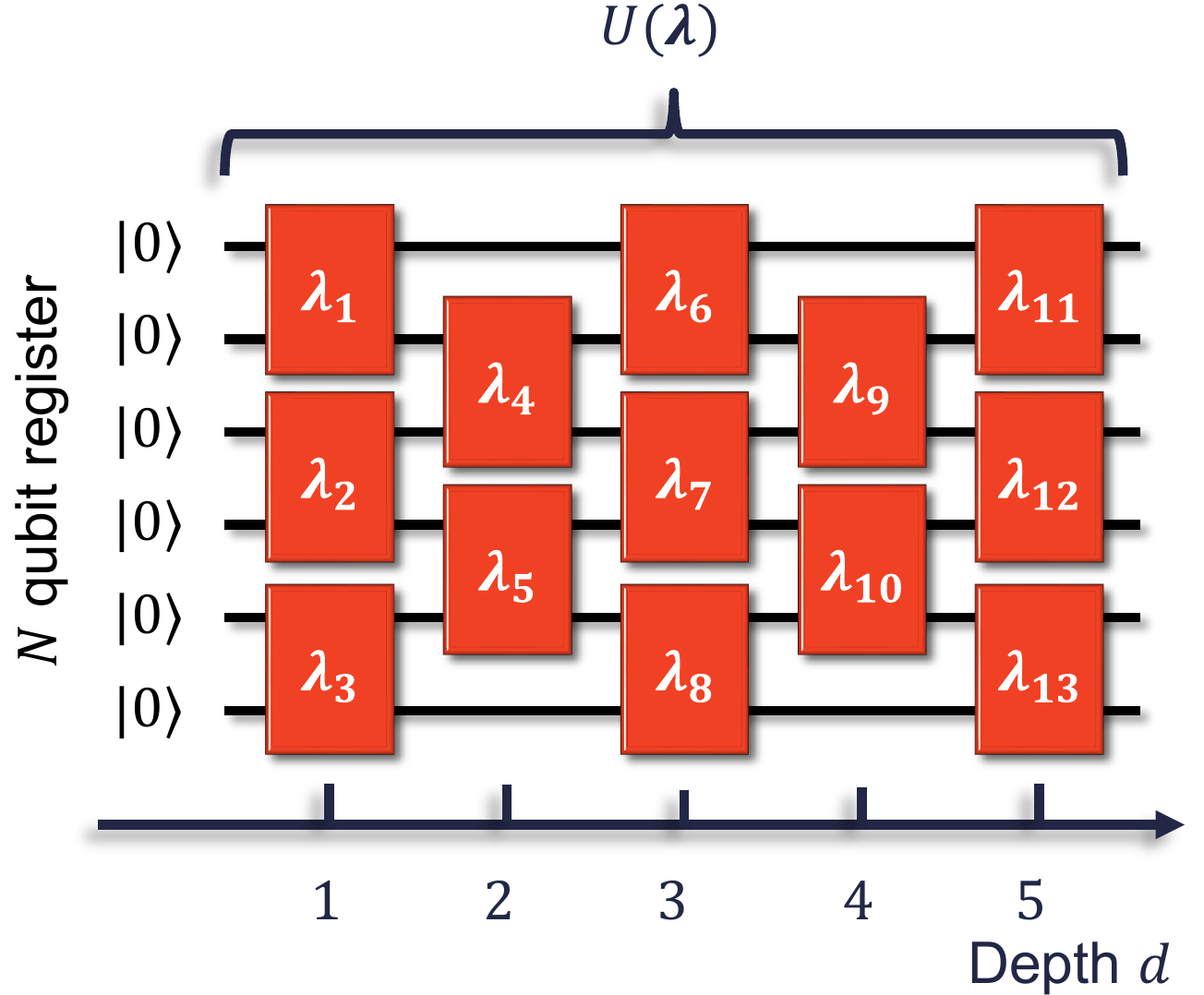}
\caption{
 Quantum ansatz of depth $d=5$. The network 
 $U(\boldsymbol \lambda)$ encodes $2^N$ function values using $N=6$ qubits. The variational parameters ${\boldsymbol \lambda}=\{\lambda_1,\lambda_2, \dots\}$ 
 determine the form of the two-qubit  gates indicated by boxes (reproduced from \cite{Lubasch_2020}).
}
\label{fig:fig4}
\end{figure}

The states $|\psi\rangle$  encoding trial functions $f$ in the multigrid encoding are created from product states initialized into $|\boldsymbol 0\rangle=|000 \dots 0\rangle$ in quantum networks shown in red in Figure \ref{fig:fig1}. The parameters defining the gates in these networks, like entanglement phase, qubit rotation axes and angles, are controlled by classical control parameters $\boldsymbol \lambda$ with an example network shown in Figure \ref{fig:fig4}. The network implements an overall unitary transformation $U({\boldsymbol \lambda})$, i.e. 
$ |\psi \rangle =U ({\boldsymbol \lambda}) |000 \dots 0\rangle$. The achievable quantum advantage of QCFD algorithms is significantly influenced by the depth $d$ of the networks required for $U({\boldsymbol \lambda})$ to be sufficiently expressive to create trial solutions $|\psi\rangle$ accurately representing the actual solutions. In general, requirements on the depth of variational circuits are not fully understood. However, because of the limited interscale correlations in CFD problems, little entanglement is needed in the trial states $|\psi\rangle$ thus limiting the required depth.

Specifically, the depth of a quantum network for generating MPS states scales with the squared bond dimension $\chi^2$. Thus, for the paradigmatic incompressible flow examples TDJ in 2D and TGV 3D discussed above, the depth of variational quantum networks representing numerically exact solutions with fidelities better than 99 \% scales with the Reynolds number like $\Rey^\gamma$ where $\gamma \approx 0$  for the 2D and 
$\gamma \approx 1.42$ for the 3D flow example. This result implies that QCFD algorithms scale more favorably than direct numerical simulations that typically go like 
$\Rey^{3K/4} \log \Rey$ where $K$ is the number of spatial dimensions of the flow. For the 2D example the scaling advantage is exponential while for the 3D example it is polynomial. Furthermore, this consideration only considered matrix-product variational trial states and thus provides a lower bound on the potentially achievable scaling advantage. Further, perhaps even exponential, improvements are possible when going beyond quantum networks that are restricted to generating MPS.

 \subsection{Programming the QNPU} 
 \label{subsec:QNPU}
 The linear operators $O_j$ acting on the trial wave functions will in general be non-unitary and can thus not directly be represented as a unitary quantum network. However, they can always be decomposed into a sum $O_j \propto \sum_{\alpha=1}^\xi U_\alpha$  of unitary operators $U_\alpha$. For many of the standard operators appearing in partial differential equations, efficient decompositions with small $\xi$ and easily implementable $U_\alpha$ are known and have been discussed in detail in the literature (see e.g. \cite{Lubasch_2020} and Section \ref{sec:BUGERS}). For non-standard operators, for instance those describing boundaries or design conditions, matrix-product operator methods can be used to create QNPU quantum networks. Because the Pauli operators form an orthogonal basis of the operator space, the operators $O_j$ can always be written as Pauli strings. The resulting representation of $O_j$ will usually possess non-optimal $\xi$ and would thus lead to quantum networks larger than necessary but can be compressed using well-known tensor-network compression techniques \cite{Schollwock2011densitymatrix} to reduce the required number of terms $\xi$. Similarly, operators representing spatially varying potentials or boundary values of flow fields can be translated into MPS making use of tensor-network tools described e.g. in \cite{Schollwock2011densitymatrix}.
 
  \subsection{Measuring cost functions and observables}
  \label{subsec:CMEASURE}
 Measuring the expectation value of the ancilla qubit in the computational basis evaluates the cost function  given by $\mathcal C=\sum_k \mathcal R\{F_k\}$, where $k$ runs over all discrete values of the function $F$ and $\mathcal R$ denotes the real part \cite{Lubasch_2020}. When carrying out the computations on quantum hardware, the expectation value is built up by averaging $\mathcal M$ individual outcomes of projective measurements on the ancilla qubit. Sampling errors scale like $1/\sqrt{\mathcal M}$ with a proportionality constant that is typically only on the order of tens to hundreds so that high accuracies can be reached with current hardware capabilities. We note that more advanced adaptive measurement schemes are being investigated and may lead to more accurate results from fewer quantum runs \cite{GarciaPerez_2021}. Furthermore, derivatives of the cost function, as required by some minimization algorithms \cite{Lubasch_2018}, can be evaluated by combining the ideas presented here with the quantum circuits discussed in \cite{Li_2017, McArdle_2019b}.
 
 The optimized variational parameters $\boldsymbol \lambda_{\rm opt}$ form a compact classical representation of the variational solution with exponentially more discrete function values that, in general, will not be computable classically from $\boldsymbol \lambda_{\rm opt}$. Hence, the strategy for evaluating desired physical quantities will be to encode them in the QNPU architecture for fixed $\boldsymbol \lambda_{\rm opt}$ as a function $A$ of the general functional form given in Eq.~(\ref{eq:FGENERATION}). Measuring the ancilla qubit will give the sum $\sum_k A_k$. For accessing individual function values $A_k$ or averages of functions over some regions of the mesh, e.g. those defined by the least significant qubits, the QNPU network needs to be augmented by controlled single qubit rotations before measuring the ancilla qubit \cite{Ekert_2002, Alves_2003}. The ability to tailor the read-out in this way and focusing measurements on specific relevant quantities is important for exploiting the quantum advantage gained during the computations when analyzing the results.
 
  \subsection{Iterative Classical Optimization}
 VQA is a highly active field of research with numerous classical optimization methods being introduced and constantly refined for quantum optimization. For instance, the quantum natural gradient method offers an improvement over simple gradient descent by calculating metrics of the cost function landscape at each optimization step \cite{Stokes_2020, Koczor_2019}. Using information sharing across multiple optimizers using Bayesian optimization, different related cost functions can be optimized in parallel and information can be shared between each optimization step to speed up convergence \cite{Self_2021}. Different widely used VQA optimization methods have been compared and studied with respect to their ability to avoid getting trapped in local minima \cite{Wierichs_2020}.
 
Simulation driven design optimization is a key development in computational aeroscience and engineering where VQA technologies naturally meet with industrial demands. Being characterized by many degrees of freedom, few cost functions and expensive design evaluations, the ideal optimizer quickly reveals a complete sensitivity map for the many parameters and advances the design. 
A gradient-based design optimization, that simultaneously solves the coupled primal/adjoint PDE system -- potentially considering additional state or geometry constraints -- in a piggy-back approach could be passed to the QNPU framework. This way the QNPU naturally addresses the optimization problem at hand. Moreover, the enormous computational power expected from quantum computing could help covering uncertainties inherent to industrial manufacturing processes using stochastic gradient approaches to identify robust design opportunities. 
\section{The Burgers' Equation}
\label{sec:BUGERS}
In this section we describe how the VQCFD method can be applied to evolve the one-dimensional Burgers’ equation in time following \cite{Lubasch_2020}. We restrict our considerations to performing an explicit, first-order Euler time integration and note that the approach can straightforwardly be extended to more advanced time-stepping methods. The nonlinear Burgers' equation for the velocity $f(x,t)$ as a function of space $x$ and time $t$ reads
\begin{equation}
\frac{\partial f}{\partial t}= \nu \frac{\partial^2 f}{\partial x^2} - f
\frac{\partial f}{\partial x} \; .
    \label{eq:PBurgers}
\end{equation}
Here, $\nu$  is the kinematic viscosity and turbulence arises when $1/\nu$ or the Reynolds number become large \cite{Bec_2007}. We discretize the spatial coordinate into equidistantly spaced discrete grid points $x_k$ and use a one-dimensional version of the multigrid method discussed in Sec. \ref{subsec:multigrid} to encode the discrete function values $f(x_k,t)$ into a quantum state $|f(t)\rangle=\lambda_0 |\psi(t)\rangle$. The Burgers'  equation then takes on the form
\begin{equation}
\frac{\partial |f(t)\rangle}{\partial t}= \nu \Delta |f(t)\rangle - D_f \nabla|f(t)\rangle \; , 
    \label{eq:VQ-Burgers1}
\end{equation}
where $\Delta$ and $\nabla$ denote discretized versions of the Laplace and Nabla operators, appropriately adapted for the multi-grid encoding, respectively. The operator $D_f$ is diagonal in coordinate space with the functions values of $f$ on its diagonal. Since the Burgers' equation does not conserve the norm of the trial function $f(x,t)$, we have introduced $\lambda_0$ as an additional variational parameter to overcome the restriction that the physical quantum state $|\psi(t)\rangle$ must always be normalized. Note that this parameter is not contained in the quantum variational network but entirely taken care of in the classical feedback loop.

The exemplary employed explicit Euler method uses the solution at time $t$ given by $|f(t)\rangle$ to compute the time-evolved solution after time step $\tau$ as $|f(t+\tau) \rangle=\big(1+\tau \; O(t)\big)|f(t)\rangle$, where $\; O(t)=\nu \Delta-D_f \nabla \; $ summarizes the discrete spatial operations. To this end we define the cost function
by the square of the residual, viz. 
\begin{equation}
    C( \, |f(t+\tau)\rangle \, )=\Big|\Big| \, |f(t+\tau)\rangle-\big( 1+\tau \, O(t) \big) \, |f(t) \rangle \, \Big|\Big|^2 \, , 
     \label{eq:VQ-Burgers2}
\end{equation}
and minimize it via varying $\lambda_0$ and the parameters ${\boldsymbol \lambda}$ of the quantum network $U({\boldsymbol \lambda})$  creating $|f(t+\tau) \rangle =\lambda_0 U({\boldsymbol \lambda})|{\boldsymbol 0} \rangle$. The state of the previous time step is written as $|f(t) \rangle=\tilde\lambda_0|\tilde \psi \rangle =\tilde\lambda_0\tilde U|
{\boldsymbol 0}\rangle$ where the parameters $\tilde \lambda_0$ and $\tilde {\boldsymbol  \lambda}$ determining $\tilde U=  U(\tilde{\boldsymbol \lambda})$ have already been fixed during the previous step. The cost function is then rewritten as a function of the variational parameters given as
\begin{equation}
    C(\lambda_0, {\boldsymbol \lambda})=
    |\lambda_0 |^2-2 \mathcal R 
    \bigg\{ 
    \lambda_0 \tilde \lambda_0^* \; 
   \langle {\boldsymbol 0}|\tilde U^\dagger \left(I+\tau \left(\nu \Delta \tilde \lambda_0 D_{\tilde \psi}  \nabla \right)\right) U({\boldsymbol \lambda})|{\boldsymbol 0} \rangle
  \bigg\}\;.
  \label{eq:OFBurgers}
\end{equation}
Here, $\mathcal R\{\cdot\}$ denotes taking the real part, $*$ the complex conjugate, $\dagger$ the hermitian conjugate, and $I$ is the identity operator.

\begin{figure}[hbt!]
\centering
\includegraphics[width=.7\textwidth]{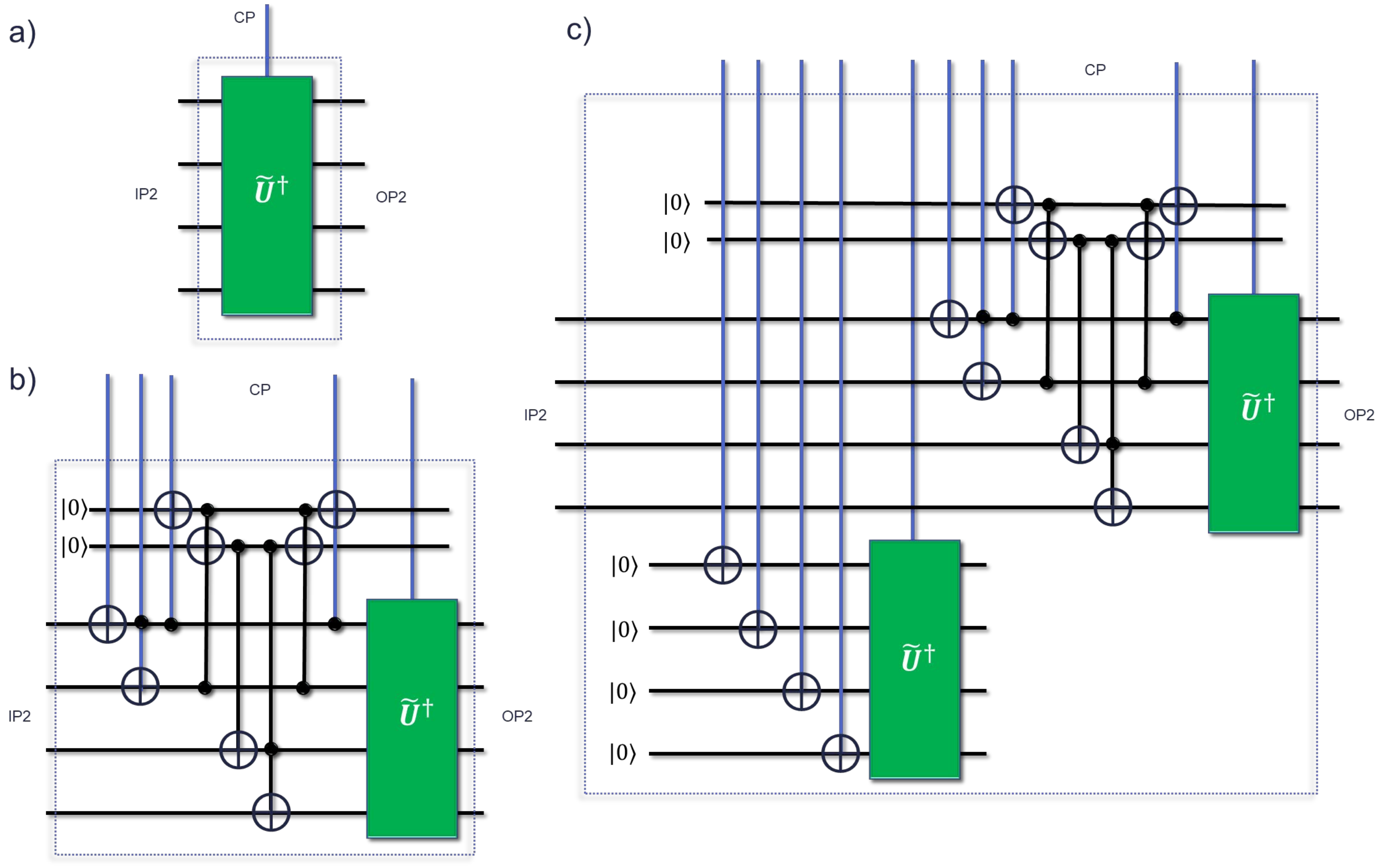}
\caption{Quantum networks required inside the QNPU to evaluate the cost function of the Burgers' equation $C(\lambda_0,\boldsymbol \lambda)$ on four qubit registers. Network a) evaluates the expectation value 
$\langle \tilde \psi | \psi \rangle =
\langle \boldsymbol 0 | \tilde U^\dagger U(\boldsymbol \lambda) | \boldsymbol 0
\rangle$, 
b) gives 
$\langle \tilde \psi | \mathcal{A}| \psi \rangle$
and c) determines the quantity 
$\langle \tilde \psi | \mathcal{A} D_{\tilde \psi}| \psi \rangle$. The operator $\mathcal{A}$ shifts the function values bitwise to calculate expectation values of derivative operators and has been adapted from the adder network \cite{Verdal_1996}. 
}
\label{fig:fig5}
\end{figure}

Figure \ref{fig:fig5} shows quantum circuits that are required for the computation of this cost function. Note that in this case only one QNPU input is utilized and we use the second controlled input IP2, i.e. $|f\rangle=|f^{(2)}\rangle$. Furthermore, the QNPU needs to be modified after every time step, as $\tilde U$ 
and hence also $|\tilde \psi\rangle$ 
change with time. As shown in Figure \ref{fig:fig5} the depth of the required quantum networks scales polynomially with the number of qubits and thus logarithmically with the size of the spatial grid. The quantum algorithm will thus be efficient if high quality variational trial functions can be created with circuits of low depths as indicated above already.

For the spatially 1D Burgers' equation one can explicitly estimate this depth. To this end one considers the initial condition $f(x,t=0)=Z \delta(x-x_0)$ with $Z$ and $x_0$ constants determining the height and position of an initial velocity hump and its evolution under the Burgers' equation \cite{Gourianov_2022, gourianov2022}. For large viscosities the solution is given by
\begin{equation}
    f(x,t)=\frac{Z}{2\sqrt{\pi \nu t}} \, e^{-\frac{(x-x_0 )^2}{4 \nu t}}.
    \label{eq:SolBurgers}
\end{equation}
When the viscosity vanishes the solution becomes triangular. For intermediate values of $\nu$ the full solution remains holomorphic in $x$. It can then be shown that all of these solutions have efficient MPS representations i.e.~are accurately approximated by products of small matrices \cite{Schollwock2011densitymatrix, Dolgov_2012, Gourianov_2022}. Since the depth of a variational quantum network to create an MPS scales with the size of its matrices,  one thus always obtains a polynomial upper bound on the depth of the variational network required for generating high-quality trial solutions \cite{Gourianov_2022}.

When attention is directed to design optimization, VQCFD could readily be employed for gradient-based, PDE-constraint 
optimization. Decoupling the optimization effort from the number of design variables usually motivates using adjoint approaches \cite{Jameson_1998},  which require solving PDEs for a set of adjoint variables that supplement the governing (primal) variables and guide the design change. As regards the Burgers' equation (\ref{eq:VQ-Burgers1}), the analogous adjoint PDE for an adjoint velocity $\hat f(x,t)$  reads
\begin{equation}
  \left( \frac{\partial {\hat f}}{\partial t} + f \frac{\partial \hat f}{\partial x} + \nu \frac{\partial^2 \hat f}{x^2}\right)  + S= 0 \, ,
    \label{eq:ADBurgers}
\end{equation}
where $S$ refers to a source term that represents objective functional influences. Eqn. (\ref{eq:ADBurgers}) can be solved along a route outlined in Eqns. (\ref{eq:VQ-Burgers1}-\ref{eq:VQ-Burgers2}).  Primal and adjoint PDEs, e.g. (\ref{eq:PBurgers}) and (\ref{eq:ADBurgers}), are usually closely linked and it is assumed that this duality also refers to the respective  trial functions. 
 For steady state problems, one could use a piggy-bag approach \cite{Hazra_2007} to iterate both the primal and the adjoint fields.  Since the VQCFD inheres features of reduced order models (ROMs), it is deemed to provide an efficient Ansatz space. To this end, an interesting aspect concerns the reversal of the adjoint transport processes. As regards the Burgers' problem, this either requires a complete storage of $f$ to compute $\hat f$, a smart check-pointing strategy which involves recomputing $f$ from a previous checkpoint \cite{Griewank_2000}, or an appropriate interpolation strategy. All options are afflicted with overhead or accuracy issues and have motivated attempts to involve improved compact descriptions of $f$. Incremental proper orthogonal decomposition (POD) methods \cite{Fareed_2018} are shown to reduce the computational cost for transient optimizations \cite{Margetis_2021} and might be suitable starting point to employ  VQCFD or classical MPS methods.

\section{VQA for Fluid Dynamics: Opportunities and Challenges}
\label{sec:VQA4CFD}
As already alluded to in the introductory section  \ref{sec:into-appl}, VQCFD can be viewed as a structure resolving ROM in which only a small portion of the potential grid space (and hence the Hilbert space for the quantum algorithm) is needed to capture the essential flow physics. Unlike the majority of current ROMs, VQCFD is not data-driven but a fully physics-based approach. The many-body physics view of turbulence structures as portrayed by their scale correlations allows the use of data compression (measured by the parameter $\chi$) to provide an accurate picture of the overall flow structures. In Ref.\ \cite{Gourianov_2022, gourianov2022} it is shown that classical simulation based on this insight with relatively low $\chi$ values  outperform under-resolved  DNS (or LES with no sub-grid scale closure) of the flow with the same computational degrees of freedom. With continuing developments of more efficient computational tensor networking routines on classical computers facilitating very high high values of $\chi$  (currently $\chi = {\cal O}(65,000)$ \cite{Ganahl2022}), comes the opportunity to simulate flows at higher Reynolds numbers. 

However, it is to be determined when VQCFD methods will be capable of competing with, or succeeding, classical CFD that currently employs up to order ${\cal O} (8,000^3)$ grid points \cite{Yeung2018}. To simulate high Reynolds number flows with a pre-specified magnitude of $\chi$, limited e.g. by the achievable depth of the quantum variational network, one may have to develop sub-structure models similar to the sub-grid scale closures in  LES \cite{Sagaut_2010}, or sub-POD/PCA models \cite{San2013}.

In non-reacting and incompressible VQCFD with small to moderate $\chi$ values, the resulting effective truncation at high wave numbers does not substantially influence the overall accuracy at large scales. Future work is needed to assess the performance of VQCFD  in more complex scenarios. In reactive flows, the effects of chemistry are dominant at very small scales, and become more significant at higher Damköhler numbers. In compressible flows, shock capturing at high Mach numbers is challenging in all discretization schemes. The same is true for resolving the fine-scale structures near  solid boundaries in turbulent boundary layers.  Future work is also recommended for VQCFD implementation for probabilistic simulations. VQCFD is then capable of tackling very high dimensional PDEs and would thus be suitable for solving the Fokker-Planck equation in RANS/PDF and LES/FDF \cite{NNGLP17}.

In summary, novel CFD approaches and methods are required to meet the future needs  in academia and industry that aim for applications beyond mere design verification. VQCFD provides a radical departure from standard ways of evolving CFD capabilities by bringing quantum computing methods into the field. Its outcomes have the potential to broadly impact the  CFD community in a variety of significant ways.  It is important, however,  to note that there are also challenges, especially in the development of quantum hardware to implement these algorithms. Although we know that a future fault-tolerant digital quantum computer would offer scaling speedups, there is a lot to be done on testing of these variational algorithms on NISQ machines to identify regimes of genuine quantum speedup over classical computing in the presence of noise. Again, this will require the combination of expertise in state-of-the art classical methods, further adaption of the algorithms to specific applications, and further input on verification and benchmarking of NISQ quantum hardware. It is thus important to continue the development of the software at pace, in order to better understand where the earliest transformative applications will occur, and on what timescale quantum hardware will be ready to enable these applications.

\section{Development of Quantum Hardware and QCFD}
\label{sec:HARDWARE}
Proof of principle simulations of the QNPU approach on IBM quantum hardware were published in 2020 \cite{Lubasch_2020}. In this work, the authors calculated the ground state of the nonlinear Schrödinger equation describing a quantum fluid and demonstrated the method’s feasibility for small and large nonlinearities. Still, those calculations were limited to small grid sizes by the available quantum hardware. The rate of improvement of quantum computing hardware is thus of paramount importance for VQCFD – and other VQA approaches -  to be able to compete with CFD on conventional computing hardware. 

Quantum hardware development is being pursued in numerous national and international research programmes. Venture capital through start-up companies and well-established large corporations like IBM and Google are also heavily invested in building quantum hardware. Using superconducting qubits, IBM promises a 1121 qubit quantum computer that would be able to hold a small number of error-corrected qubits by 2023 \cite{Cho_2020}. Ion trap experiments have already demonstrated fault-tolerant control of an error-corrected qubit in small-scale devices \cite{Egan_2021}. In addition, there are numerous hardware and software start-up companies that collaborate and/or compete with these efforts and add unique innovative and transformational aspects to the field. For instance, alternative all-optical quantum computing platforms are pursued by Xanadu \cite{Arrazola_2021} and PsiQuantum, who aim to build fully scalable and error corrected quantum computers \cite{psiquantum_2022}. There is similarly a great deal of excitement about the potential for neutral atom quantum computing \cite{Bluvstein_2022, Graham_2022}.

If only one of the current hardware efforts succeeds and meets their targets, then a quantum advantage will be achievable in industrially relevant problems, with VQCFD being a particularly promising early application, this decade already. In order to realize this, though, it will be important to prepare for the availability of this new generation of quantum hardware now, and further develop VQCFD and other algorithms for tackling engineering simulations in the real world, as noted in the previous section.

\section{Discussion and Outlook}
\label{sec:OUTLOOK}
In this article we have focused on the example of VQCFD to convey the type of approach that needs to be taken to utilizing quantum computing for applications in aeroscience. There are several ways in which this is representative of the general opportunities and challenges in applying quantum computing to real-world applications.

Firstly, we need to take an approach that is different from  conventional algorithms. In the case of VQCFD, the core of how we use the quantum hardware (storing a variational state that we cannot directly access) is very different to conventional CFD algorithms, and needs to be in order to take advantage of quantum hardware. Quantum hardware can usually not directly be used to implement conventional algorithms with any speedup at all, and so finding the right new approach for a specific problem will be the critical step in opening new application areas.

Secondly, because we need to find new approaches, it is vital to bring together experts in quantum algorithms with experts in classical CFD, in order to understand the breadth of opportunities to write the relevant problems in different mathematical language, and to understand what needs to be done to create algorithms that usefully go beyond the capabilities of existing classical algorithms for the given application area. For VQCFD there is already a strong connection between engineers working on CFD and experts in quantum computing. This will become only more important as we extend the example calculations for VQCFD to real-world problems, and begin to understand where it will demonstrate clear advantages over existing techniques.

In addition, although our main focus is on producing algorithms for rapidly developing quantum hardware, there are already positive spin-offs from the developments here in the form of quantum-inspired classical algorithms. Because approaching problems on a quantum computer requires finding new approaches that generally have not previously been attempted on a classical computer, there is the possibility that we find algorithms that beat existing classical techniques. Quantum inspired computing has already become important in the development of optimization tools. Here, by using tensor network approaches, VQCFD techniques have been mapped onto new CFD algorithms on classical computers, which may provide advantages in solving certain classes of problems even while the appropriate quantum hardware is still under development. 

Quantum computing has the potential to make a large impact on aeroscience and engineering. But there are challenges in developing applications, and these are best addressed by strong interdisciplinary engagement. We hope to see applications in the coming decade, but it is important already now to work on algorithms that can be tailored to specific computational problems and to specific developing hardware platforms. This will allow the joined-up communities to identify the most promising future applications, and the timescales on which the hardware will reach the required scale and level of precision to realize these in real-world applications.

\section*{Acknowledgements}
D.J. and T.R. acknowledge support by the European Union’s Horizon Programme (HORIZON-CL4-2021-DIGITAL-EMERGING-02-10) Grant Agreement 101080085 QCFD. D.J. and A.D. were supported by EPSRC Programme Grant DesOEQ (EP/P009565/1). D.J. acknowledges support from AFOSR grant FA8655-22-1-7027 and by the Excellence Cluster ‘The Hamburg Centre for Ultrafast Imaging—Structure, Dynamics and Control of Matter at the Atomic Scale’ of the Deutsche Forschungsgemeinschaft. A.D. acknowledges support from AFOSR grant number FA9550-18-1-0064.

\bibliography{./references.bib, references_givi.bib, cfd_givi.bib}

\end{document}